\documentclass[usenatbib]{mn2e}

\input psfig.sty

\title[Gas near a QSO at z=4.28]{Fluorescent Lyman-alpha Emission from Gas Near a
QSO at Redshift 4.28}
\author[Francis \& McDonnell]{
Paul J. Francis$^{1,2}$\thanks{E-mail:
pfrancis@mso.anu.edu.au (PJF)} and
Sunsanee McDonnell$^{1}$ \\
$^{1}$Research School of Astronomy and Astrophysics, the Australian
National University, Canberra 0200, Australia\\
$^{2}$Joint Appointment with the Department of Physics,
Faculty of Science, the Australian National University
}
\begin{document}

\maketitle

\label{firstpage}

\begin{abstract}

We use integral field spectroscopy with the Gemini North Telescope to detect probable 
fluorescent Ly$\alpha$ emission from gas lying close
to the luminous QSO PSS 2155+1358 at redshift 4.28. The emission is most likely coming 
not from primordial gas, but from a multi-phase, chemically enriched cloud of gas lying
about 50 kpc from the QSO. It appears to be associated with a highly ionised associated absorber
seen in the QSO spectrum. With the exception of this gas cloud, the environment of
the QSO is remarkably free of neutral hydrogen. We also marginally detect Ly$\alpha$ emission
from a foreground sub-Damped-Ly$\alpha$ absorption-line system.

\end{abstract}


\section{Introduction}

Flows of gas play a crucial role in galaxy formation \ -- \ both the infalling
primordial gas from which galaxies ultimately form, and outflows which enrich the
intergalactic medium and regulate galaxy formation. Much attention has recently been
placed on one possible way of directly observing this gas: Ly$\alpha$ fluorescence.
The idea, originally proposed by \citet{hog87} is that the plentiful Ultraviolet (UV) 
radiation in
the high redshift universe will be absorbed by any neutral gas, and some fraction of
the incident ionising photons will be re-radiated as Ly$\alpha$ photons 
\citep{gou96,bun98,fra01,fra04,can05}.

This fluorescent Ly$\alpha$ emission will be brightest where the UV radiation is
strongest: close to luminous QSOs. \citet{hai01} and \citet{ala02} predicted that 
QSOs with sufficiently high redshifts should be surrounded by halos of Ly$\alpha$ emission, as the
UV flux from the QSO causes the still infalling primordial gas to fluoresce. The predicted
intensity of this flux is, however, very dependent upon assumptions on how the gas is
clumped.

The observational picture is currently confusing. It has long been established that many
high redshift radio sources, both QSOs and radio galaxies, have bright Ly$\alpha$ emission
coming from around them \citep[e.g.][]{hec91,bre92,hu96,ven02}, but this may be caused by 
the radio jets, winds or cooling flows, rather than fluorescence.
Luminous extended Ly$\alpha$ nebulae  (``blobs'') are fairly common at high redshifts, but
not obviously associated with QSOs: they may be caused by cooling flows, superwinds or
photoionisation by concealed AGNs \citep[e.g..][and refs therein]{fur03,mat04,col06}.

There are less data available for radio-quiet QSOs. Damped
Ly$\alpha$ absorption-line systems that lie close to the QSO redshift have been detected in 
possible fluorescent emission in a few cases \citep{mol93,mol98}. More recently, probable 
fluorescent emission was seen from another absorption system, but this time the absorber was
380 kpc away, in front of a different QSO \citep{ade06}. \citet{fra04} failed to detect any
fluorescent emission around a QSO, and indeed claimed that the presence of the QSO was suppressing
even the normal background population of Ly$\alpha$ emitting galaxies. On the other hand,
\citet{wei04} detected Ly$\alpha$ fuzz close to a $z = 3$ QSO which they modeled as a spherical
infalling halo of primordial gas. \citet{bun03} found spectacularly bright Ly$\alpha$ fuzz, with 
a high velocity dispersion, around a QSO at redshift 4.46. \citet{bar03} suggest that evidence for
infalling primordial gas around high redshift QSOs can be seen in the detailed profiles of the QSO
emission line.

It is thus becoming clear that fluorescent Ly$\alpha$ can be detected with current telescope
facilities, but on the basis of the extremely small number statistics available to date, its
properties appear to be heterogeneous. A larger sample size would clearly be useful. In this paper
we contribute to this effort, searching for fluorescent emission near the brightest accessible
QSO with a redshift above four: PSS 2155+1358, at redshift 4.28.

Previous detections of fluorescent Ly$\alpha$ have used either long-slit spectroscopy
\citep[e.g.][]{bun03} or narrow-band imaging \citep[eg.][]{fra04}. Long-slit spectroscopy
has the disadvantage that the slit may not be at the right orientation to catch any emission.
Narrow-band imaging is only possible if the QSO redshift is known with high precision, which is
not usually the case. In this paper we try a different approach: integral field unit (IFU)
spectroscopy.

We assume a cosmology with $H_0 = 70{\rm km\ s}^{-1}{\rm Mpc}^{-1}$, $\Omega_{\rm matter} = 0.3$
and $\Omega_{\Lambda} = 0.7$ throughout this paper.

\section{Observations and Data Reduction\label{observe}}

The QSO PSS 2155+1358 was observed with the integral field unit of 
the Gemini Multi-Object Spectrograph (GMOS) on the Gemini North Telescope. 
The B600 grating was used in two-slit mode, giving a spectral resolution of $R = 1688$, over the 
5$\times 7\arcsec$ IFU field of view. The $r\_G0303$ filter was used
to limit the wavelength range to 562 -- 698nm and hence prevent spectra from the
two slits from overlapping. A central wavelength of 625nm was used, 
with some dithering in the spectral direction between frames to eliminate chip
gaps.

Observations were obtained in queue mode. Observations were taken on 2004 August 14, 
September 11 and September 14, in clear weather conditions and with seeing varying between 
0.7 and 0.9 arcsec. A total exposure time of 6 hours was obtained.

The data were reduced using the Gemini IRAF (Image Reduction and Analysis) package. Standard
settings were used to flat field and bias subtract the frames, extract the spectra,
wavelength calibrate and sky subtract the spectra, and assemble them into a data cube.
There was a problem with the wavelength calibration, due to a (now rectified) issue with
the grating mounts. As a result, the absolute wavelength calibration of each frame, based on arc 
lines, was out by a small amount: different frames had to be aligned using
absorption-lines in the spectra before co-addition. We used sky lines to make sure that the wavelength 
calibration was at least self-consistent between all the fibres in a give slit. The final co-added 
data cube was put on 
a correct wavelength scale by comparing the absorption line centroids with those measured in
an independent higher resolution spectrum (\S~\ref{absorb}). The product was a data cube with 0.2 
arcsec spatial pixels and 0.092nm spectral pixels.

Sky subtraction used a separate 5$\times 3.5\arcsec$ sky region, one arcmin from the
QSO. This subtraction (as performed by the Gemini IRAF package) left a flat residual,
probably due to scattered light: this residual was weakly wavelength dependent but differed
considerably between the two slits. This led to a wavelength dependent step in the 
background level between the left and right hand sides of any image slice through the data cube.
As the QSO sat exactly on the boundary between the two halves, this would have been
disastrous for point spread function subtraction if it had not been corrected. An IRAF 
script was written to perform this correction. For each spatial slice through the data cube,
the top and bottom three pixels in each column were medianed, and this median subtracted from
the whole column.

\subsection{Searching for diffuse line emission\label{compact}}

We first searched for diffuse line emission which might fill the whole IFU field of view, 
as would be seen if the QSO was embedded, for example, in a giant Ly$\alpha$ blob.
For this measurement, the scattered light correction mentioned in the previous section was not
used.

In each $5 \times 7\arcsec$ spatial slice, we averaged the two ends: the first and last
$5 \times 1\arcsec$ regions, which were free of QSO light. The average of these
two regions was plotted, looking for narrow spikes.

The largest spikes had an amplitude (average, over both $5 \times 1\arcsec$ regions) of $1.2 \times 
10^{-18}{\rm erg\ cm}^{-2}{\rm s}^{-1}{\rm arcsec}^{-2}$ over a three spectral pixel (1300
${\rm km\ s}^{-1}$) velocity range. This is a conservative upper limit: over the majority 
of the spectrum (away from the rare strong sky lines) the limit is a factor of three better.

\subsection{Searching for compact emission}

We next searched the data cube for any compact emission-line sources. This was done both
with and without subtraction of the QSO point spread function.

Point spread function (PSF) subtraction was done using an IRAF script as follows. For each
spectral slice through the data cube, a separate point spread function was calculated. It was formed 
from the median of all the spectra slice images in the velocity ranges $-4300$:$-1300$ and 
$+1300$:$+4300 {\rm km\ s}^{-1}$. Images in which the QSO flux was very low due to an absorption line
were excluded from this median. The PSF image thus constructed was scaled to have the correct
flux in the central $ 0.6 \times 0.6\arcsec$ region of the QSO, and subtracted from the spectral slice 
image. This procedure worked very well in general, giving PSF subtraction errors typically
smaller than the background noise. Several other PSF subtraction techniques were used, but gave
poorer results. These other techniques were, however, used as a check on any objects discovered.

Unfortunately, a sub-set of image slices had errors considerably above the
background noise, for a variety of reasons:

\begin{itemize}

\item Wavelength shifts between the two halves of the image. It turned out that the wavelength 
solution for the two halves of the image (deriving from the two GMOS slits) was different, albeit
only at the sub-pixel level. This introduced artifacts in the PSF subtracted image where the
gradient of the QSO spectrum was steep, typically at the edges of narrow absorption lines. 
Unfortunately our spectral sampling was not sufficient to fully remove it. It produced, however,
an easily recognised characteristic symmetrical left-right pattern, and only occurred at the
edge of strong narrow absorption lines. 

\item Chip gaps. The GMOS camera uses a three-CCD mosaic, which produced gaps in the wavelength
coverage of any individual spectrum. We observed at four different central wavelengths, ensuring
that we had at least three quarters of our data at any given wavelength. As the seeing differed 
between the different nights' data, however, the seeing width in the co-added data-cube 
varied depending which nights data
were used at a given wavelength. This led to strong residuals in the PSF subtraction near the edges of
any of the regions affected by a chip gap. Once again, this produced a very characteristic pattern
(a positive or negative ring in the subtracted image) and it was easy to check that such rings 
were indeed produced by a chip gap.

\item Other glitches. There were occasional glitches away from the QSO position, due to
cosmic ray hits, defects in the CCDs and problems in the spectrum extraction software. These were
mercifully rare, and typically easy to recognise because they affected either a region smaller than
the PSF, or a whole row or column in the data. A disadvantage of fibre IFU designs such as GMOS
is that the mapping between pixels in the raw data and those in the final data cube is very
complex and not one-to-one, making it extremely time-consuming checking the origin of glitches such
as these.

\end{itemize}

For all these reasons, it did not prove possible to do an automated search for line emission. 
Instead, the whole cube was eyeballed independently by the two authors, and lists of candidate
emission-line objects assembled. These candidates were then checked in detail to eliminate the
various glitches. Errors were estimated from the size of the largest candidate objects for which 
it was not possible to confidently decide whether they were glitches or not. This error limit 
typically a factor of $\sim 2$ greater than the ($3 \sigma$) sky noise limit.

\begin{figure*}
 \psfig{figure=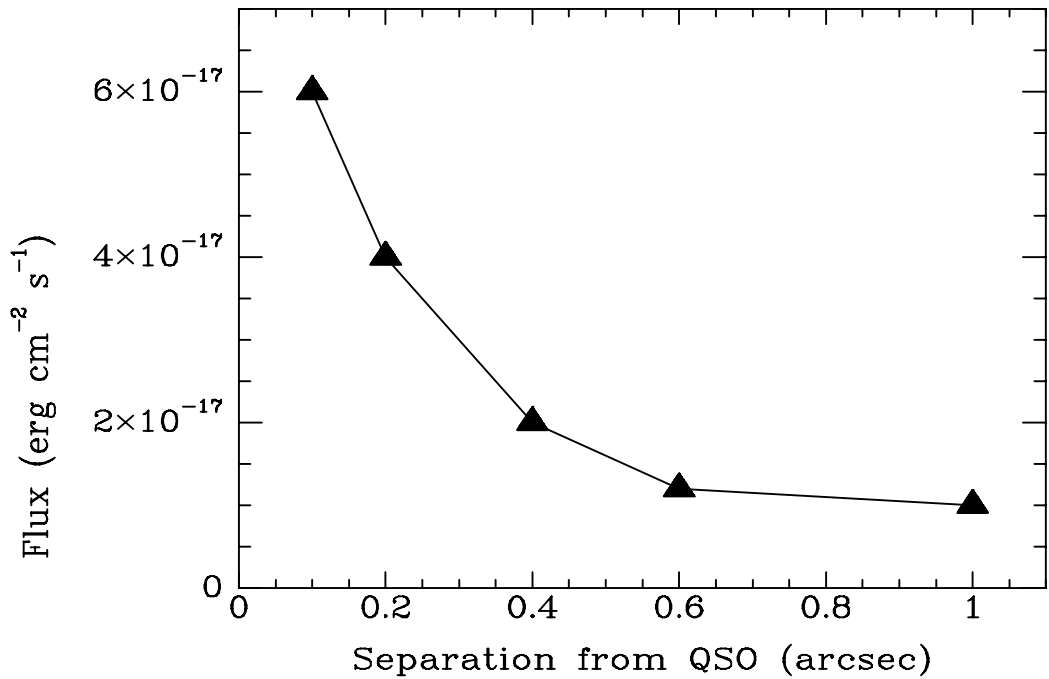}
  \caption{
  Detection limits for spectrally unresolved point source line emitters, as a function
  of distance from the QSO. This curve was determined near the peak of the QSO Ly$\alpha$ emission
  line: sensitivities will be better a wavelengths where the QSO is fainter.\label{limits}
  }
\end{figure*}

Our detection limit was estimated by inserting artificial point-sources into various
image slices, carrying out the point spread function subtraction and seeing if they
were obviously seen, and easily distinguished from the various artifacts. As shown in 
Fig~\ref{limits}, our sensitivity to point sources is a strong function of how far they are 
in projection from the QSO. 

\section{Results}

\begin{figure*}
 \psfig{figure=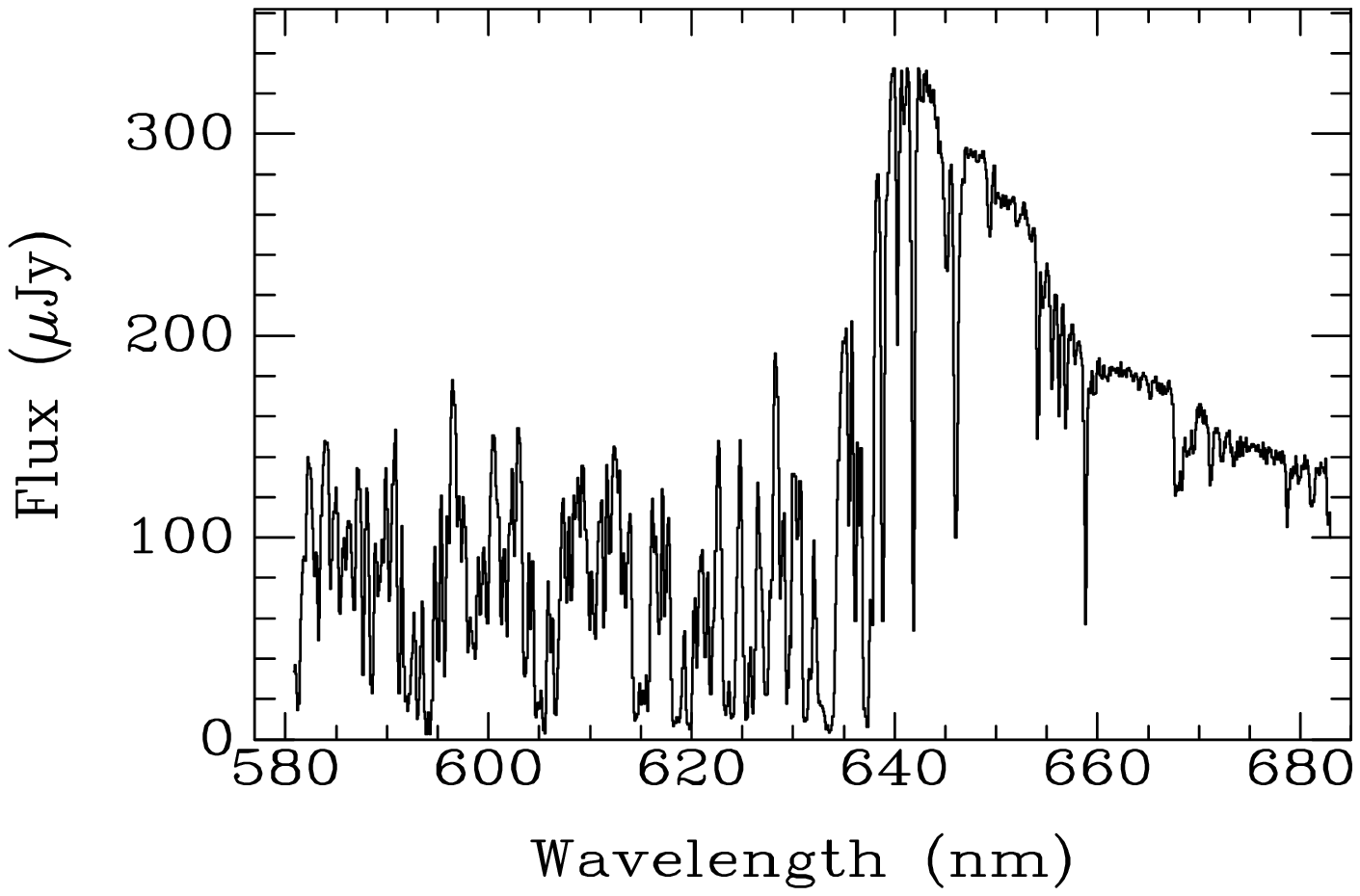}
  \caption{
  Spectrum of QSO PSS 2155+1358, extracted from the GMOS data cube.\label{qsospec}
  }
\end{figure*}

The QSO spectrum is shown in Fig~\ref{qsospec}, and shows a strong Ly$\alpha$ line peaking
at around 642 nm, and many absorption lines. The observed QSO flux is 181 $\mu$Jy at 660nm.

Only one significant Ly$\alpha$ emitter was discovered, at 641.97nm
(We defer discussion of a second, marginally significant emitter to \S~\ref{marginal}).
This emission-feature can be clearly seen in each night's data independently, so we regard the
detection as secure.

\begin{figure*}
 \psfig{figure=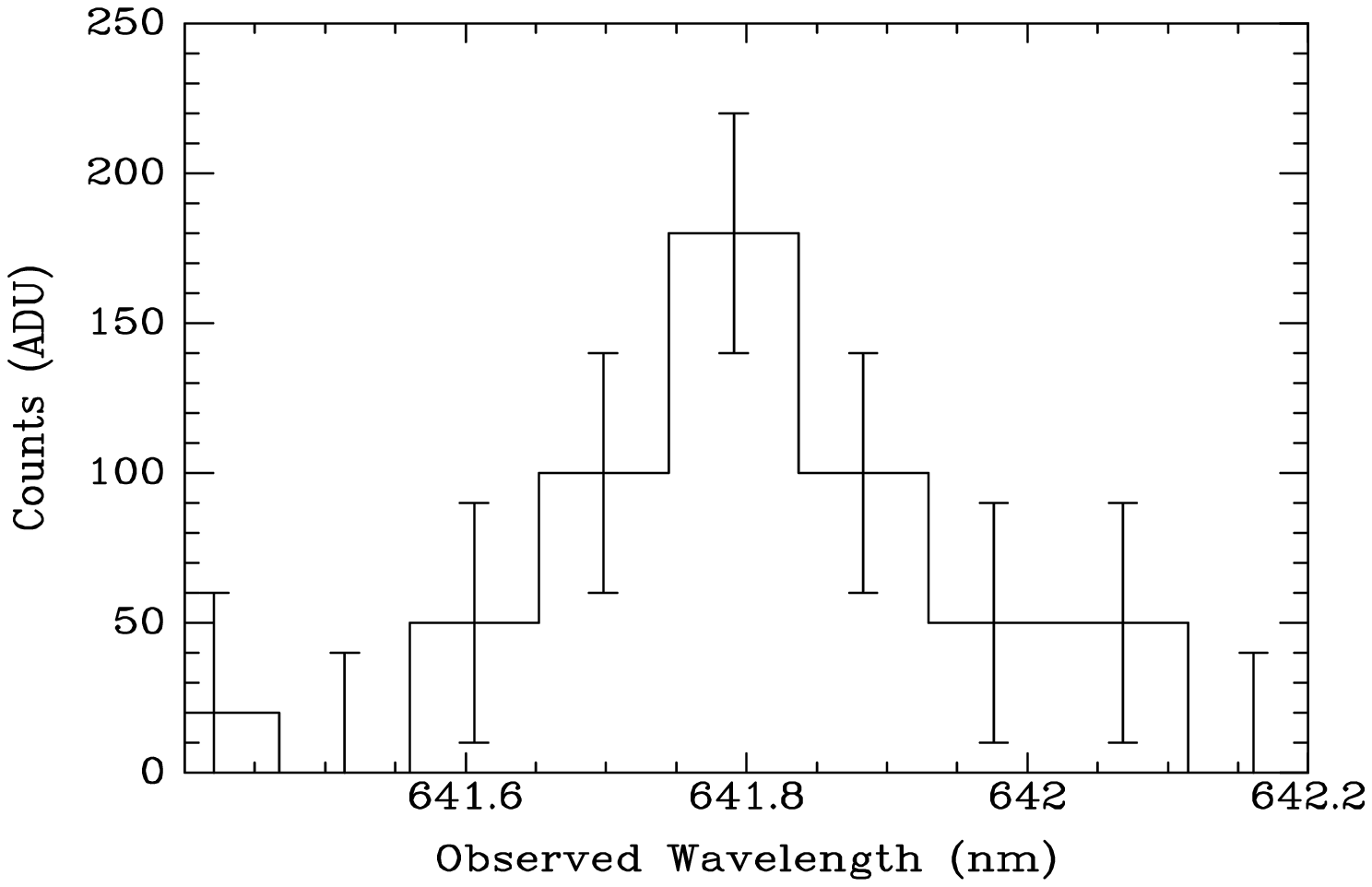}
  \caption{
  Spectrum of the candidate Ly$\alpha$ emission line at z=4.2793.\label{profile}
  }
\end{figure*}

The 641.97nm line emission comes from $0.8\arcsec$ east of the QSO. It is spectrally unresolved
(Fig~\ref{profile}) with a line width (full width at half maximum height, FWHM) of 
$\sim 0.3$nm ($140 {\rm km\ s}^{-1}$). A flux of 
$1.5 \times 10^{-17}{\rm erg\ cm}^{-2}{\rm s}^{-1}$ was measured by point-spread function fitting
of the QSO PSF subtracted image. Note that this is a lower limit on the true flux, as any flux 
overlapping the centre of the QSO image will have been removed by the QSO PSF subtraction. 
No continuum emission is detected at this location: this allows us to place a lower limit of 
$\sim 10$nm (observed frame) on the equivalent width of the line. 

\begin{figure*}
 \psfig{figure=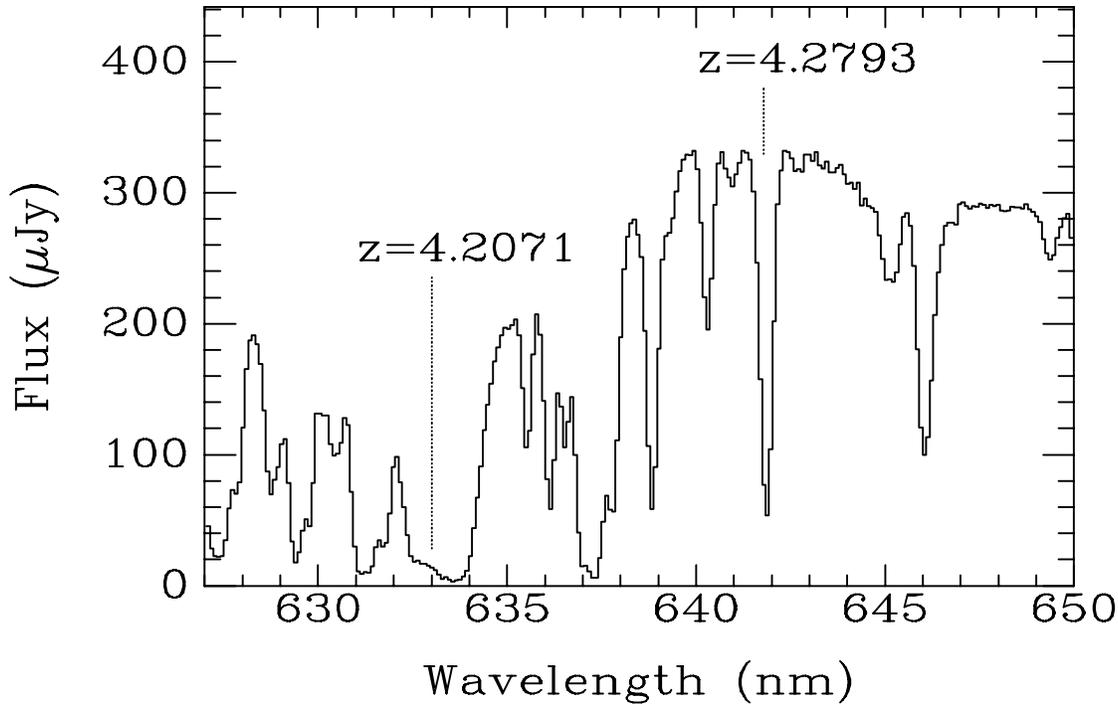}
  \caption{
  Close-up of the spectrum of QSO PSS 2155+1358, showing the wavelength of the
  main candidate Ly$\alpha$ emitter (z=4.2793), and of the second, more marginal
  detection (z=4.2071).\label{qsospec_close}
  }
\end{figure*}

The line lies at the peak of the
broad QSO Ly$\alpha$ emission-line, at almost the same wavelength as a strong narrow Ly$\alpha$
absorption-line system (Fig~\ref{qsospec_close}). The emission peak is offset by 
${45 \pm 30\rm km\ s}^{-1}$ ($1 \sigma$ uncertainty) 
to the blue of the absorption-line centroid. The emitter is not an artefact of the wavelength
calibration problem at the edge of strong absorption-lines, noted in \S~\ref{compact}: it lies on
the flat part of the QSO spectrum near the base of the absorber, where this artefact is not
important.

\begin{figure*}
 \psfig{figure=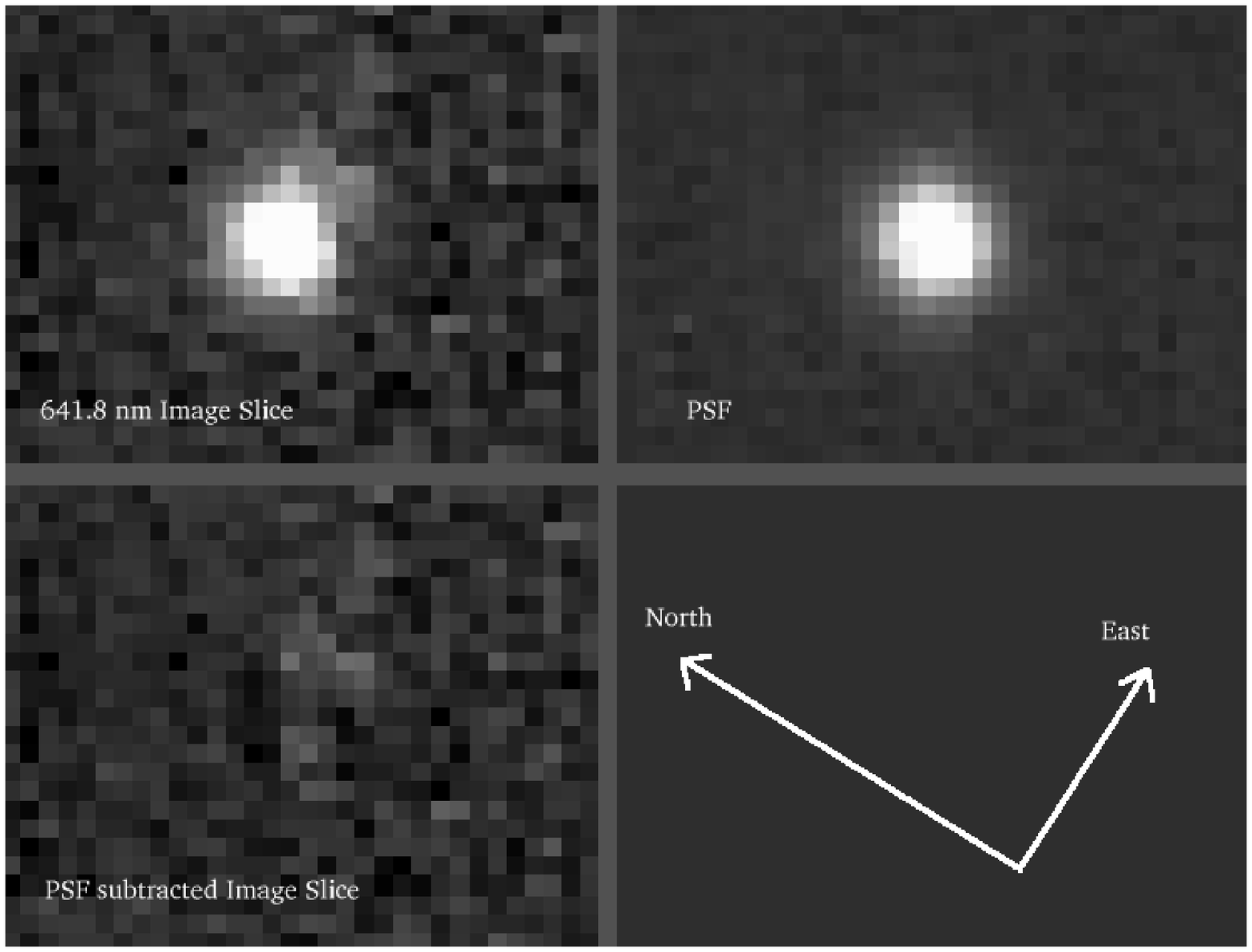}
  \caption{
The candidate Ly$\alpha$ emitter at redshift 4.2793. The top left panel shows a 
single spatial slice through the data cube at 630.00nm wavelength. The point spread function
generated from other slices near this wavelength is
shown in the top right panel. The bottom left panel shows the PSF subtracted 
image. The emitter is above and to the right of the QSO.
Each image is $7\arcsec$ wide and $5\arcsec$ high. Upwards in the images 
corresponds to $50^{\circ}$ east of north on the sky. 
  \label{grid660_label}
  }
\end{figure*}

\subsection{Absorption Line}

\label{absorb}

What are the properties of this absorption-line, which lies at the same redshift as the
Ly$\alpha$ emitter?
Celine P\'eroux kindly made available to us a spectrum of QSO PSS 2155+1358 obtained with the
Ultra-Violet Echelle Spectrograph (UVES) on the Very Large Telescope (VLT). This spectrum 
is described in \citet{des03}. This spectrum was used to investigate the properties
of the absorption-line at the same wavelength as the z=4.2793 emitter.

The line is clearly seen in multiple Lyman-series lines, allowing an accurate 
neutral hydrogen column density of $4.68 \times 10^{15} {\rm cm}^{-2}$ to be measured. 
The line width parameter $b = 32 {\rm km\ s}^{-1}$.

\begin{figure*}
 \psfig{figure=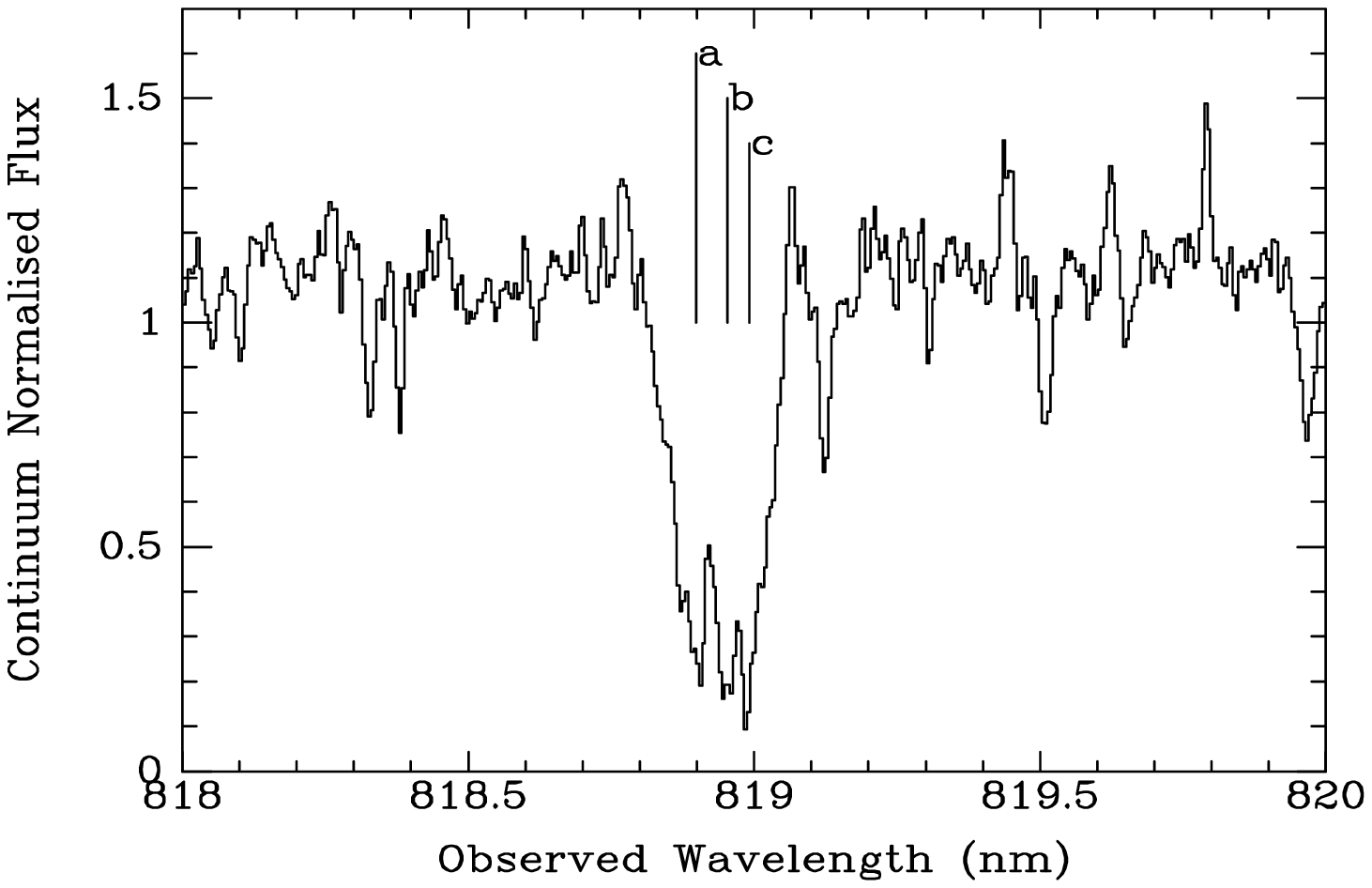}
  \caption{
  The UVES spectrum of the CIV (155.077 nm) absorption-line, showing the three 
  components.\label{plotc4}
  }
\end{figure*}

The line is also detected in N~V and C~IV, where it breaks up into at least three sub-components 
each with a velocity width of no greater than 10${\rm km\ s}^{-1}$ (Fig~\ref{plotc4}), spread 
over  a 40${\rm km\ s}^{-1}$ velocity range. We fit the absorption with a three-sub-component
model: component $a$ is at a vacuum heliocentric redshift of 
z=4.2806, component $b$ at z=4.2809 and component $c$ at z=4.2812.

Column densities were measured interactively using the XVOIGT program \citep{mar95}
The C~IV column densities are 6.3, 9.1 and $8.5 \times 10^{13} {\rm cm}^{-2}$ for components
$a$, $b$ and $c$ respectively. For N~V, component $a$ has a column density $6.31 \times 10^{13} 
{\rm cm}^{-2}$, while the combined column density of components $b$ and $c$ is $1.16 \times 10^{14} 
{\rm cm}^{-2}$. Upper limits on the column 
density of each individual component in other transitions are given in Table~\ref{ulims}.

\begin{table}
 \centering
 \begin{minipage}{80mm}
  \caption{Upper limits on the absorption column density of each absorbing component.\label{ulims}}
  \begin{tabular}{@{}lc@{}}
  \hline
   Ion     &  $\log_{10}(Column Density)$, ${\rm cm}^{-2}$ \\
 \hline
    N~III & 13.8 \\
    N~II  & 12.5 \\
    N~I   & 14.5 \\
    C~III & 13.3 \\
    C~II  & 12.8 \\
    C~I   & 12.5 \\
    Si~IV & 12.5 \\
    Si~II & 11.9 \\
    Si~I  & 13.3 \\
    Fe~III & 14.3 \\
    Fe~II & 13.2 \\
    Al~II & 11.5 \\
    Ni~II & 12.5 \\
    S~I   & 13.5 \\
    O~I   & 13.5 \\
\hline
\end{tabular}
\end{minipage}
\end{table}

\section{Discussion}

\subsection{Is the line Ly$\alpha$?}

Is this emission line due to Ly$\alpha$ emission at redshift 4.2793, or could it be emission
from a foreground galaxy, in a different line? The detected line is too narrow to be [O~II] (372.7nm),
as we would have easily resolved the doublet. The main possibilities are thus H$\beta$ and
the [O~III] doublet. If it were any of these lines, we might have expected to see one of the others.
The expected wavelengths for all the various possibilities were checked and nothing was seen.

We can put a conservative lower limit on the equivalent width of the line of 10nm. This is quite
normal for Ly$\alpha$ emitting galaxies, but would be
extremely high for a foreground galaxy \citep[eg.][]{fra04b}.

We therefore conclude that the line is most likely Ly$\alpha$. Further evidence for this hypothesis
comes from the probability argument in \S~\ref{redshift}.

\subsection{What is the QSO's Redshift?}

To work out whether the Ly$\alpha$ emitter we saw is associated with the QSO, we need to know the QSO
redshift. \citet{per01} estimated the QSO redshift by fitting Gaussians to the peak of the
Ly$\alpha$, Si~IV/O~IV (140nm) and C~IV (154.8nm) lines. Their data consisted of a spectrum with a relatively 
low resolution and signal-to-noise ratio. The results 
were discrepant: z=4.285 from Ly$\alpha$, 4.269 from Si~IV/O~IV and a much lower value of 4.216 
from C~IV. The discrepancy was unsurprising, as all three lines were strongly affected by absorption-line
systems.

We used the UVES spectrum to better determine the redshift. Its higher resolution and 
signal-to-noise ratio allowed us to clearly see the continuum between the absorption lines. From 
this spectrum, we see that the Ly$\alpha$ emission line really does have quite a sharp peak, and that
the downturn below 639nm is real and not due to absorption. Using this peak wavelength, we get a
redshift estimate of $4.28 \pm 0.02$. Measuring a redshift from the metal lines proved to be
harder, as both have extremely broad, flat profiles, without a well defined peak. Our best estimate is
$4.26 \pm 0.04$ for Si~IV/O~IV, and $4.25 \pm 0.04$ for C~IV.

The results are thus still discrepant, albeit all overlapping within their respective error bars.
\citet{fra92} showed that CIV is systematically blue-shifted in radio-quiet QSOs, and that this
blueshift is strongest when the line is broadest and has a flat top. The narrow core of Ly$\alpha$,
if present, seems to be a more reliable redshift indicator, provided you have enough spectral 
resolution to measure the true continuum shape through the Ly$\alpha$ forest absorption.

Another clue to the redshift comes from a series of associated C~IV absorption lines, covering
a range of redshift, with the highest at z=4.2793. Unless this gas is infalling rapidly into the QSO,
this places a lower limit on the true redshift.

Taken together, we thus prefer the higher redshift measured in Ly$\alpha$: 
$z_{\rm QSO} = 4.28 \pm 0.02$.

\subsection{Is the Emitter at the QSO's Redshift?}

\label{redshift}

In this section we argue that the odds of detecting a Ly$\alpha$ emitter by chance close in 
both redshift and projected distance to the QSO are small. We therefore conclude that the
emitter we saw is indeed physically close to the QSO.

\citet{hu98} measure the surface density of Ly$\alpha$ emitters at z=4.5 down to a flux limit 
close to ours.
Using their surface density, we would only have expected to find $\sim 0.08$ emitters in our whole
data cube. The odds of seeing one by chance within one projected arcsec of the QSO is $< 1$\%.
And the odds of finding one that is also within the redshift range $4.28 \pm 0.02$ is 
$< 0.07$\%. We can also do the probability calculation
internally to our data, by noting that no Ly$\alpha$ emitters were found in the outer regions of
our data cube (20 square arcsec, $\Delta z = 0.82$) while one was seen within one arcsec
of the QSO in the redshift range $4.28 \pm 0.02$. The ratio of volumes is thus 
128:1.

It is thus quite unlikely that we would have found our one source so close to the QSO by
chance. We therefore conclude that the probability of finding a Ly$\alpha$ emitter is enhanced close
to the QSO, and that the QSO and emitter really lie close to each other.

\subsection{Are the Emitter and Absorber Connected?}

The wavelength of our Ly$\alpha$ emitter lies within $45 {\rm km\ s}^{-1}$ ($0.1$ nm)
of the wavelength of an associated absorption-line system in the QSO spectrum. Is this coincidence, or could the 
two be physically connected? 

How likely is such a coincidence? Consider a random Ly$\alpha$ emitter, which lay at a redshift
within our uncertainty on the redshift of the QSO ($4.28 \pm 0.02$), and which we would thus 
consider to be associated with the QSO. There are two strong absorption-lines within this
redshift interval (Fig~\ref{qsospec_close}). The odds of a random emitter lying within $0.1$ nm
of one of these absorbers is 8\%. This assumes no wavelength dependence in our sensitivity
limit. In practice, we are more sensitive where the QSO spectrum is weaker, but our emitter is strong
enough to have been detected regardless of the QSO flux at that wavelength.

This number is small, but not small enough to rule out the possibility that the redshift match
between absorber and emitter is coincidence. We conclude that while it is quite likely that 
the absorber and emitter are connected, this has not been proved.

\subsection{Physical Properties of the Gas}

We have concluded that the Ly$\alpha$ emitter at $z = 4.2793$ lies close to the QSO, and
may well be connected with the gas producing the absorption line $45 {\rm km\ s}^{-1}$ further
to the red. What can we learn about the physical properties of this absorber?

For our adopted cosmology, the Ly$\alpha$ emitting cloud is spectrally unresolved (FWHM 
$< 150 {\rm km\ s}^{-1}$), spatially either marginally resolved or unresolved 
(size $<  5 \times 5$kpc), and has a luminosity of $2.6 \times 10^{42}{\rm erg\ s}^{-1}$: more
if we've over-done the PSF subtraction. It lies at a projected distance of 5kpc from the QSO

Let us first assume that the cloud is $4 \times 4$ kpc in size, and optically thick at the 
Lyman-limit. Following the method in \citet{fra04}, we can estimate the fluorescent Ly$\alpha$ 
emission that the incident QSO flux will induce in it, as a function of its (proper) distance from
the QSO. If it were $\sim 5$kpc from the QSO, its predicted luminosity would be $\sim 400$ times
greater than that observed. 

There are several ways to explain its faintness:

\begin{itemize}

\item The cloud, while seen close in projection to the QSO, is actually $\sim 100$kpc in-front of 
or behind the QSO, and thus exposed to less ionizing radiation.

\item The cloud is much smaller than the seeing disk: less than 1 kpc on a side.

\item The cloud, while large, is mostly optically thin, with only a small filling factor of
optically thick gas reprocessing the QSO flux.

\item Dust or optical depth effects suppress the Ly$\alpha$ flux.

\item The QSO could have been much fainter a few thousand years ago: due to light travel time,
the cloud luminosity reflects the QSO luminosity in the past.

\end{itemize}

If, however, we assume that the emitting gas is connected with the absorption-line system, we can
deduce much more about its physical conditions, and rule out several of these possibilities.

Firstly, if the emitting and absorbing gas come from different parts of the same cloud, 
that gives us a cloud size of at least 5kpc. Secondly, the absorption consists of three sub-clumps,
spread over $\sim 40{\rm km\ s}^{-1}$. If the cloud really had a very small filling factor of
these absorbing sub-clumps, the odds of finding three along the QSO sight-line would have been
small. It is therefore likely that the filling factor of sub-clumps is such that most sight-lines
would intersect at least one. Thirdly, the measured neutral hydrogen column is too small to
make the clouds optically thick in Ly$\alpha$, and the fact that we see the background QSO through 
the clouds implies that little dust is present.

We modeled the gas using version 06.02.09 of Gary Ferland's Cloudy photoionisation code
\citep{fer98}. The absorbing sub-clumps were modeled as plane-parallel slabs of
constant density gas, exposed to continuum emission with a typical QSO spectrum, normalised to
the observed luminosity of PSS 2155+1358. 

We place a lower limit on the ionisation parameter $U > 0.1$. This limit comes from
our observed lower limit on the ratio of the C~IV to C~III column density, and is, to first
order, independent of the density. No upper limit on $U$ can be placed. The gas is therefore
highly ionised: the fraction of hydrogen that is in the neutral state is $< 2 \times 10^{-5}$.
While the measured neutral column density is only $4.7 \times 10^{15} {\rm cm}^{-2}$, the 
total ionised hydrogen column is more like that of a damped Lyman-$\alpha$ absorption-line system. 
For $U = 0.1$, we get a good fit to the observed column densities of both nitrogen and carbon if the
metallicity is 6\% (-1.2 dex) of solar. If the ionisation parameter is larger, we need carbon to
be overabundant with respect to nitrogen - a trend which has been seen before in QSO spectra
\citep[eg.][]{ham99}.

Each sub-clump reprocesses around 15\% of incident ionising photons into Ly$\alpha$ photons. If 
most sight-lines through the cloud intersect a sub-clump (as suggested by the presence of three along the
QSO sight-line), this means that the cloud must be around 50 Kpc from the QSO to have the luminosity
observed. If it is further, an additional ionisation source would be required.

How big are the sub-clumps? If they are 50kpc from the QSO, and $U=0.1$, the inferred gas density 
needed to give this ionisation parameter is $\sim 3 \times 10^2{\rm cm}^{-3}$. The total gas column 
is $\sim 10^{20}{\rm cm}^{-2}$, giving a sub-clump thickness of $3 \times 10^{17}{\rm cm}$.
If $U \sim 1$, the inferred density drops by a factor of ten and the
ionised column increases by an order of magnitude, giving sub-clumps a hundred times as thick.

Thus if the absorbing gas is associated with the Ly$\alpha$ emitter, we're not looking at the host
galaxy of the QSO. Instead, we are looking at an isolated cloud of metal-enriched gas many
tens of kpc away, perhaps a satellite galaxy of the QSO host. If, however, the absorber is 
not connected with the emission, we have fewer constraints, and could be looking at gas 
closer in, the emission from which is suppressed by one of the mechanisms listed in the
bullet points above.

If the emitter really is $\sim 50$ kpc from the QSO, as required if it is connected with the
absorption line, then it is a little surprising that we see it at a projected separation of only 
$0.8\arcsec$ from the QSO. 50kpc in the plane of the sky would correspond to a separation of 8 
arcsec. If emitters are distributed randomly within 50 kpc of the QSO, 28\% would lie within
our data cube, and only 2.5\% would lie within one projected arcsec of the QSO. Thus if we saw 
anything, the odds of it being within $1\arcsec$ are $\sim 10$\%. This is not unlikely enough to
rule anything out, but is curious and perhaps provides weak evidence that the emitter is
closer to the QSO and hence not connected with the absorber.

\section{The Marginal Emitter at Redshift 4.2071}

\label{marginal}

\begin{figure*}
 \psfig{figure=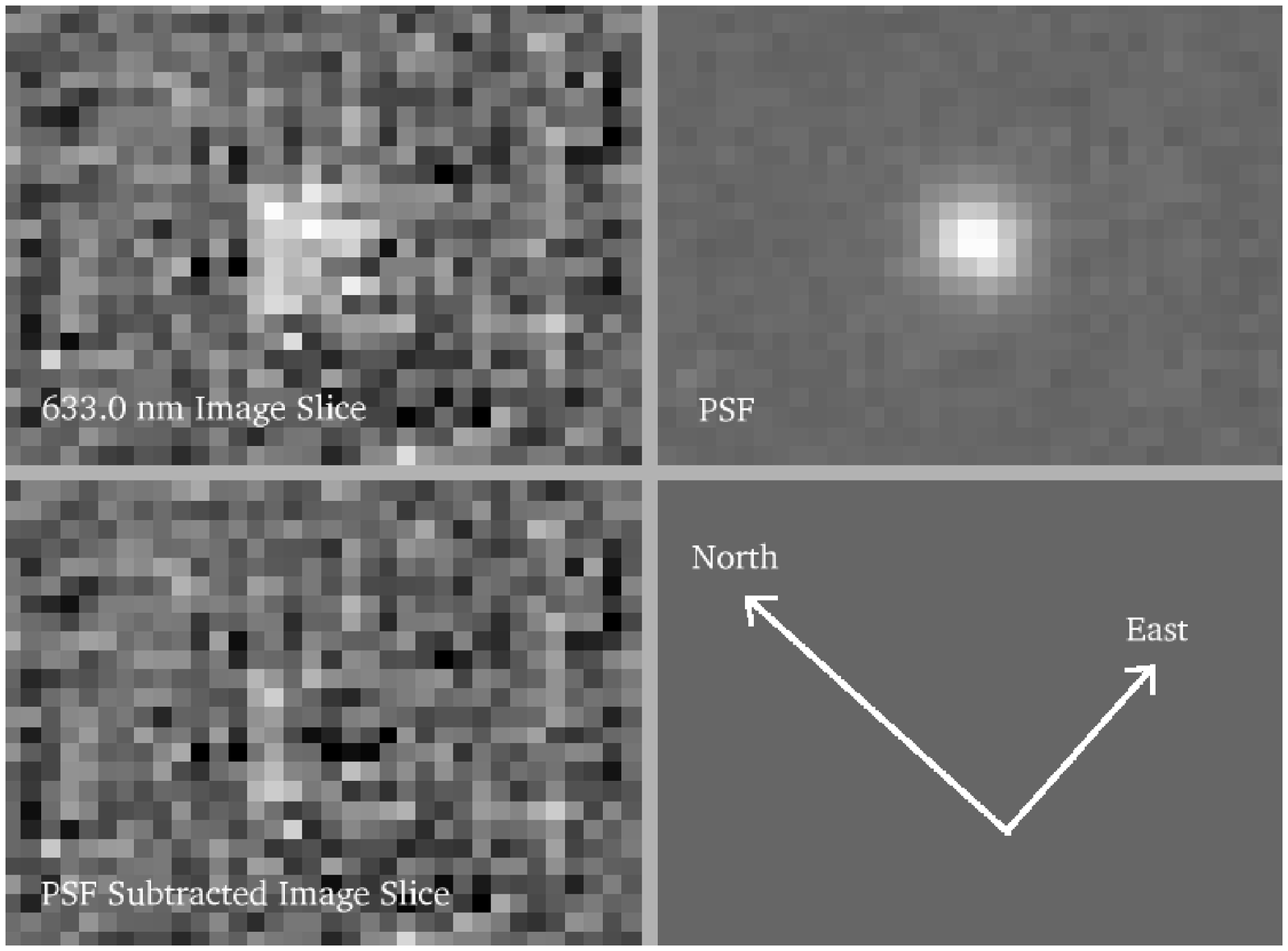}
  \caption{
The candidate Ly$\alpha$ emitter at reshift 4.2071. The top left panel shows a 
single spatial slice through the data cube at 630.00nm wavelength. The point spread function
generated from other slices near this wavelength is
shown in the top right panel. The bottom left panel shows the PSF subtracted 
image. The candidate emitter lies below and to the left of the QSO.
Each image is $7\arcsec$ wide and $5\arcsec$ high. Upwards in the images 
corresponds to $50^{\circ}$ east of north on the sky. 
  \label{grid565_label}
}
\end{figure*}

In addition to the main emission-line at z=4.2793, a second, more marginal line is
seen at z=4.2071 (Fig~\ref{grid565_label}). This is located $\sim 0.8\arcsec$ west of the QSO, and 
has a flux of about $1.0 \times 10^{-17}{\rm erg\ cm}^{-2}{\rm s}^{-1}$. It too is 
spectrally unresolved. Formally, it is significant at about the $6 \sigma$ level, but it is
too faint to see clearly in the data from individual nights, and hence could be some sort 
of glitch in the data.

This emitter lies in the blue wing of a sub-Damped-Lyman$\alpha$ absorption-line system
in the QSO spectrum (Fig~\ref{qsospec_close}). This absorption system was studied by 
\citet{des03} and consists of two components, at redshifts of 4.212229 and 4.212628. 
If the existence of this emitter is confirmed, and if it is indeed associated with
this nearby absorber, it would join the
small number of such absorption-line systems seen in emission 
\citep[eg.][and refs therein]{war01,mol02,wea03,kul06}.

\section{Conclusions}

Perhaps our most striking conclusion is how little neutral gas lies close to QSO PSS 2155+1358.
We detect one isolated cloud, which is probably $\sim 50$kpc from the QSO. But there is no extended
Ly$\alpha$ nebula, as seen around radio galaxies, radio-loud QSOs and in Ly$\alpha$ blobs. The 
emission we do detect is much fainter than that seen by \citet{wei04} and \citet{bun03}. Clearly
the properties of gas around high redshift QSOs are diverse.

\section*{Acknowledgments}

Based on observations obtained at the Gemini Observatory, which is operated by the
Association of Universities for Research in Astronomy, Inc., under a cooperative agreement
with the NSF on behalf of the Gemini partnership: the National Science Foundation (United
States), the Particle Physics and Astronomy Research Council (United Kingdom), the
National Research Council (Canada), CONICYT (Chile), the Australian Research Council
(Australia), CNPq (Brazil) and CONICET (Argentina). The program number was GN-2004B-Q-21.
Data from the European Southern Observatory's Very Large Telescope were also used. We'd like
to thank C\'eline Peroux for making available the reduced UVES spectrum of PSS 2155+1358.0

\bsp

\label{lastpage}

\end{document}